\documentclass[twocolumn,showpacs,preprintnumbers,amsmath,amssymb]{revtex4}

\usepackage{graphicx}
\usepackage{bm}

\usepackage{amsfonts}

\begin{document}
\title{Two-dimensional multisolitons and azimuthons in Bose-Einstein condensates with
attraction}
\author{Volodymyr M. Lashkin}
\email{vlashkin@kinr.kiev.ua}
 \affiliation{Institute for Nuclear
Research, Pr. Nauki 47, Kiev 03680, Ukraine}

\date{\today}

\begin{abstract}

We present spatially localized nonrotating and rotating
(azimuthon) multisolitons in the two-dimensional (2D)
("pancake-shaped configuration") Bose-Einstein condensate (BEC)
with attractive interaction. By means of a linear stability
analysis, we investigate the stability of these structures and
show that rotating dipole solitons are stable provided that the
number of atoms is small enough. The results were confirmed by
direct numerical simulations of the 2D Gross-Pitaevskii equation.
\end{abstract}


\maketitle

Localized coherent structures, such as fundamental solitons,
vortices, nonrotating and rotating multisolitons are universal
objects which appear in many nonlinear physical systems
\cite{Kivshar1}, and, in particular, in Bose-Einstein condensates
(BEC's). Stability of these nonlinear structures is one of the
most important questions because of its direct connection with the
possibility of experimental observation of solitons and vortices.

Detailed investigations of the stability of localized vortices in
an effectively two-dimensional (2D) trapped BEC with a negative
scattering length (attractive interaction) were performed in Ref.~
\cite{Malomed} and later extended to the three-dimensional case
\cite{Malomed2} (see also Ref.~\cite{Saito}). While vortex
solitons in attractive BEC are strongly unstable in free space,
the presence of the trapping potential results in existence of
stable vortices provided that the number of particles does not
exceed a threshold value \cite{Malomed,Saito,Berge}.

Recently, a novel class of 2D spatially localized vortices with a
spatially modulated phase, the so called azimuthons, was
introduced in Ref.~\cite{Kivshar2}. Azimuthons represent
intermediate states between the radially symmetric vortices and
rotating soliton clusters. In contrast to the linear vortex phase,
the phase of the azimuthon is a staircaselike nonlinear function
of the polar angle. Various kinds of azimuthons have been shown to
be stable in media with a nonlocal nonlinear response
\cite{Lopez1,Lopez2,Skupin,We2,We3}.

The aim of this Brief Report is to present nonrotating
multisoliton (in particular, dipole and quadrupole) and rotating
multisoliton (azimuthon) structures in the 2D BEC with attraction
and study their stability by a linear stability analysis. We show
that, in the presence of a confining potential, stable azimuthons
also exist for a medium with a local cubic attractive
nonlinearity. Results of the linear stability analysis were
confirmed by direct numerical simulations of the azimuthon
dynamics.

We consider a condensate which is loaded in an axisymmetric with
respect to the $(x,y)$ plane harmonic trap, and tightly confined
in the $z$ direction. The dynamics of the condensate is described
by the Gross-Pitaevskii equation (GPE)
\begin{gather}
i\hbar\frac{\partial \Psi}{\partial
t}=\left\{-\frac{\hbar^{2}}{2m}\Delta+\frac{m}{2}[\Omega_{r}^{2}(x^{2}+y^{2})+
\Omega_{z}^{2}z^{2}]\right.
\nonumber \\
\left. +\frac{4\pi\hbar^{2}a}{m}|\Psi|^{2}\right\}\Psi,
\label{eq1}
\end{gather}
where $\Psi(\mathbf{r},t)$ is the condensate wave function,
$a\gtrless 0$ is the $s$-wave scattering length. We assume that
the axial confinement frequency $\Omega_{z}$ is much larger than
the radial one $\Omega_{r}$ (the pancake configuration). Then, the
3D equation (\ref{eq1}) can be reduced to an effective GPE in two
dimensions \cite{Malomed} (for a detailed discussion of the
applicability of the 2D approximation see Ref. \cite{Malomed}).
The standard reduction procedure \cite{Salasnich, Louis} leads to
the following 2D equation
\begin{equation}
\label{main} i\frac{\partial\psi}{\partial
t}=-\Delta_{\perp}\psi+\frac{\Omega_{0}^{2}}{4}(x^{2}+y^{2})\psi-\sigma|\psi|^{2}\psi,
\end{equation}
where appropriate dimensionless units are used, and
$\Delta_{\perp}=\partial/\partial x^{2}+\partial/\partial y^{2}$,
$\sigma=\pm1$, where the $+(-)$ sign corresponds to attractive
(repulsive) contact interaction.

Equation (\ref{main}) conserves the 2D norm (the normalized number
of particles)
\begin{equation}
N=\int |\psi|^{2}dxdy,
\end{equation}
and energy
\begin{equation}
E=\int
\left\{|\nabla_{\perp}\psi|^{2}+\frac{\Omega_{0}^{2}}{4}(x^{2}+y^{2})|\psi|^{2}
-\frac{\sigma}{2}|\psi|^{4}\right\}dxdy
\end{equation}

Stationary solutions of Eq. (\ref{main}) in the form
\begin{equation}
\psi(x,y,t)=\varphi (x,y)\exp(-i\mu t),
\end{equation}
where $\mu$ is the chemical potential, resolve the variational
problem $\delta S=0$ for the functional
\begin{equation}
\label{S}
 S=E-\mu N.
\end{equation}
Following Ref.~\cite{Berge}, let us consider the trial function
$\varphi=A\phi_{a}(r/a,\theta)$, where $\phi_{a}$ is some test
function, $A$ and $a$ is the amplitude and characteristic width of
the stationary state, respectively. Then, the functional $S$ takes
the form
\begin{equation}
\label{func}
 S(a,A)=\alpha A^{2}-\beta
A^{4}a^{2}+\frac{\Omega_{0}^{2}}{4}\gamma A^{2}a^{4}-\mu\delta
A^{2}a^{2},
\end{equation}
where the integral coefficients are
\begin{gather}
\label{coe1} \alpha=\int
|\nabla_{\vec{\xi}}\,\phi_{a}|^{2}d\vec{\xi},\quad
\beta=\frac{\sigma}{2}\int |\phi_{a}|^{4}d\vec{\xi},\\
\label{coe2} \delta=\int |\phi_{a}|^{2}d\vec{\xi},\quad
\gamma=\int \xi^{2}|\phi_{a}|^{2}d\vec{\xi}.
\end{gather}

The Euler-Lagrange equations $\partial S/\partial a=0$ and
$\partial S /\partial A=0$ for the functional Eq. (\ref{func})
give expressions for the width $a$ and amplitude $A$ \cite{Berge}
\begin{equation}
\label{width}
a^{2}(\mu)=\frac{2\sqrt{\mu^{2}\delta^{2}+3\alpha\gamma\Omega_{0}^{2}}+
2\mu\delta}{3\Omega_{0}^{2}\gamma}
\end{equation}
\begin{equation}
\label{A}
A^{2}(\mu)=\frac{\Omega_{0}^{2}a^{2}\gamma-2\mu\delta}{2\beta}
\end{equation}
The 2D norm $N$ and energy $E$ then read
\begin{equation}
\label{N} N(\mu)=A^{2}(\mu)a^{2}(\mu)\delta=\frac{\delta
a^{2}(\mu)}{2\beta}(\Omega_{0}^{2}a^{2}\gamma-2\mu\delta),
\end{equation}
\begin{equation}
\label{E}
E(\mu)=\frac{A^{2}(\mu)a^{4}(\mu)}{2}\Omega_{0}^{2}\gamma .
\end{equation}
In the following, we consider attractive short-range interactions
and set $\sigma=1$. It then follows from Eq. (\ref{A}) that
solutions exist provided that the chemical potential $\mu$ does
not exceed a critical value $\mu^{\ast}$ \cite{Berge},
\begin{equation}
\label{mucr}
\mu\leq\mu^{\ast}=\frac{\Omega_{0}}{\delta}\sqrt{\alpha\gamma}.
\end{equation}
To proceed further, we take a test function $\phi_{a}$ in the form
\begin{equation}
\label{trial}
\phi_{a}(\vec{\xi})=\xi^{m}L_{n}^{(m)}(\xi^{2})e^{-\xi^{2}/2}(\cos
m\theta+i\,p\sin m\theta),
\end{equation}
where $m$ is an integer, $\theta$ is the azimuthal angle,
$0\leqslant p\leqslant 1$, and $L_{n}^{(m)}(x)$ is the $n$th (i.
e. with $n$ zeros) generalized Laguerre polynomial. The parameter
$p$ determines the modulation depth of the soliton intensity. Note
that the case $p=0$ corresponds to the nonrotating multisolitons
(e. g. $m=1$ to a dipole, $m=2$ to a quadrupole etc.), while the
opposite case $p=1$ corresponds to the radially symmetric
vortices. The intermediate case $0<p<1$ corresponds to the
rotating azimuthons. Since the vortices and azimuthons have a
nontrivial phase, they carry out the nonzero $z$-component of the
angular momentum
\begin{equation}
M_{z}= \mathrm{Im}\,\int
\left[\psi^{\ast}(\mathbf{r}\times\nabla_{\perp}\psi)\right]_{z}
dx\,dy,
\end{equation}
which can be expressed as
\begin{equation}
M_{z}(\mu)=\frac{2pm}{(p^{2}+1)}N(\mu).
\end{equation}
When $n\neq 0$, the ansatz Eq. (\ref{trial}) represents a rather
exotic structure with $n$ nodes.

Inserting Eq. (\ref{trial}) into Eqs. (\ref{coe1}) and
(\ref{coe2}), one can determine the coefficients $\alpha$,
$\beta$, $\gamma$ and $\delta$. For the nodeless case ($n=0$) we
have
\begin{gather}
\label{coef0}
\alpha_{0}^{(m)}=\gamma_{0}^{(m)}=\frac{\pi}{2}(p^{2}+1)(m+1)!,\\
\beta_{0}^{(m)}=\frac{\pi}{2^{m+5}}(3p^{4}+2p^{2}+3)(2m)!,\\
\delta_{0}^{(m)}=\frac{\pi}{2}(p^{2}+1)m!. \label{coef00}
\end{gather}
For the structures with one node ($n=1$), one can obtain
\begin{gather}
\label{coef1}
\alpha_{1}^{(m)}=\frac{\pi(m+3)}{2}(p^{2}+1)(m+1)!,\\
\beta_{1}^{(m)}=\frac{\pi(3m^{2}+5m+2)}{2^{2m+7}}(3p^{4}+2p^{2}+3)(2m)!,\\
\delta_{1}^{(m)}=\frac{\pi}{2}(p^{2}+1)(m+1)!,\\
\gamma_{1}^{(m)}=\frac{\pi}{2(m+2)}(p^{2}+1)(m+3)!. \label{coef11}
\end{gather}
Substituting Eqs. (\ref{coef0})--(\ref{coef00}) or Eqs.
(\ref{coef1})--(\ref{coef11}) into Eqs. (\ref{width}) and
(\ref{A}), we get the width and amplitude of the corresponding
nonlinear structures. In what follows we restrict ourselves to the
case of the nodeless states ($n=0$) and, in addition, set
$\Omega_{0}=2$.

\begin{figure}
\includegraphics[width=3.4in]{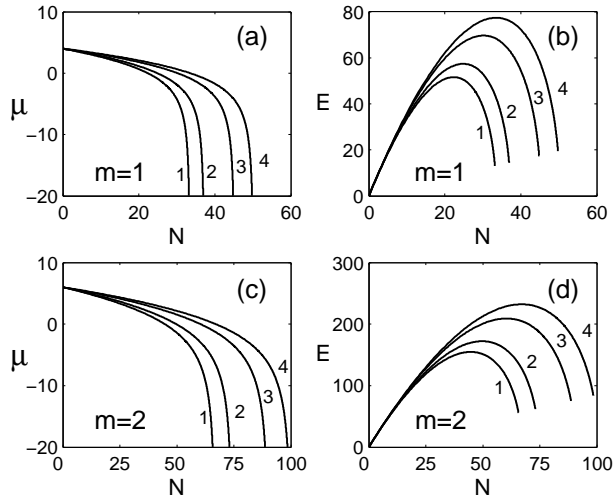}
\caption{\label{fig1} (a) Chemical potential $\mu$ and (b) energy
$E$ versus normalized number of atoms $N$ for variational nodeless
($n=0$) solutions Eq. (\ref{trial}) with
 $p=0$ (curve $1$), $p=0.3$ (curve $2$), $p=0.6$ (curve $3$), and $p=1$ (curve $4$) for $m=1$;
 (c) and (d) the same for $m=2$. }
\end{figure}

The dependences $\mu (N)$ and $E(N)$ obtained from Eqs. (\ref{N})
and (\ref{E}) for different values $p$ are plotted in
Fig.~\ref{fig1} for $m=1$ and $m=2$. The curve $4$ corresponding
radially symmetric vortices $p=1$ coincides with that obtained in
Ref.~\cite{Malomed,Berge}. It follows from Eq. (\ref{mucr}) that
the critical value for the chemical potential is
$\mu^{\ast}=2(m+1)$ and does not depend on $p$. The asymptotic
values $N_{max}(\mu=-\infty)$, which determine the (formal)
collapse threshold, decrease with decreasing $p$. Results of the
variational analysis are found to be in good agreement with
numerical simulations (see below).

\begin{figure}
\includegraphics[width=3.4in]{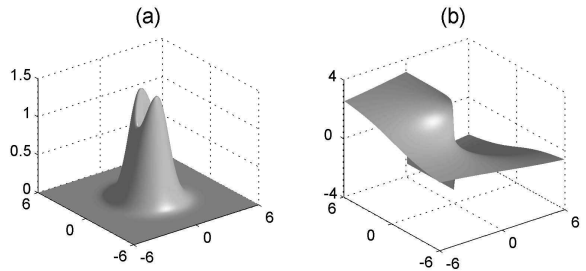}
\caption{\label{fig2} (a) Amplitude $|\varphi|$ and (b) phase
$\arg\varphi$ distributions of the azimuthon with two intensity
peaks ($m=1$) and $p=0.7$, $\mu=2.7$. }
\end{figure}

Generally speaking, using the relaxation technique similar to one
described in Ref. \cite{Petviashvili} and choosing an appropriate
initial guess, one can find numerically vortex and azimuthon
solutions of Eq. (\ref{main}) on Cartesian grid. Under this, the
parameter $p$ (modulational depth), which is similar to the one in
Eq. (\ref{trial}), can be introduced in the following way:
\begin{equation}
p=\max|\mathrm{Im}\,\Psi|/\max|\mathrm{Re}\,\Psi|.
\end{equation}
However, the choice of initial guess (to achieve convergence) is
extremely difficult and time consuming, and, moreover, we were not
able to find azimuthon solutions with arbitrary $p$. Instead, we
use an approximate but much simpler variational approach and
introduce the following ansatz in polar coordinates $(r,\theta)$
\cite{Lopez1}
\begin{equation}
\label{trial1} \psi(\mathbf{r},t)=U(r)(\cos m\theta+i\,p\sin
m\theta)e^{-i\mu t},
\end{equation}
where $U(r)$ is a real function. Inserting the ansatz
(\ref{trial1}) into the action Eq. (\ref{S}), integrating over
$\theta$, but keeping an arbitrary dependence $U(r)$, one can then
obtain the corresponding Euler-Lagrange equation,
\begin{equation}
\label{vareq}
\frac{d^{2}U}{dr^{2}}+\frac{1}{r}\frac{dU}{dr}+\left(\mu-r^{2}-\frac{m^{2}}{r^{2}}\right)U
+f(p)U^{3}=0,
\end{equation}
where
\begin{equation}
f(p)=\frac{3p^{4}+2p^{2}+3}{4(p^{2}+1)}.
\end{equation}
Equation (\ref{vareq}) was solved numerically with boundary
conditions $U\rightarrow r^{|m|}$ at $r\rightarrow 0$, and
$U\rightarrow 0$ at $r\rightarrow\infty$. In Fig. \ref{fig2} we
demonstrate an example of the azimuthon with two intensity peaks
(i. e. with the topological charge $m=1$), $p=0.7$, and $\mu=2.7$.
For fixed chemical potential $\mu$ and integer $m$, there is a
family of azimuthon solutions with different $p$. Note that the
ansatz (\ref{trial1}) represents only particular class of the
azimuthons. More general form of the azimuthon solutions, which
are characterized by two independent integer numbers (number of
peaks is generally independent on the topological charge $m$), was
introduced in Ref.~\cite{Kivshar2}.

\begin{figure}
\includegraphics[width=3.4in]{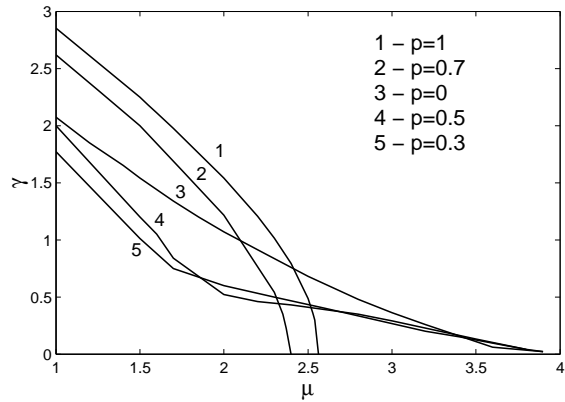}
\caption{\label{fig3} The growth rates $\gamma$ as functions of
the chemical potential $\mu$ for $m=1$ and different $p$.}
\end{figure}

To study the stability of the stationary solutions, we represent
the wave function in the form
\begin{equation}
\label{per}
\psi(\mathbf{r},t)=[\varphi_{0}(\mathbf{r})+\varepsilon(\mathbf{r},t)]e^{-i\mu
t},
\end{equation}
where the stationary solution $\varphi_{0}$ is perturbed by a
small perturbation $\varepsilon$. As usual, one could then take
\begin{equation}
\varepsilon(\mathbf{r},t)=\varphi_{+}(\mathbf{r})e^{i\omega
t}+\varphi_{-}^{\ast}(\mathbf{r})e^{-i\omega t},
\end{equation}
and consider the corresponding eigenvalue problem, but, in our
case with $p\neq 1$, this approach turns out to be ineffective.
Indeed, the linearization of Eq. (\ref{main}) around $\varphi_{0}$
in $\varepsilon$ leads to the eigenvalue problem
\begin{gather}
\hat{T}\varphi_{+}+\varphi_{0}^{2}\varphi_{-}=\omega\varphi_{+},
\label{eigen1} \\
-\hat{T}\varphi_{-}-\varphi_{0}^{\ast\,2}\varphi_{+}=\omega\varphi_{-},
\label{eigen2}
\end{gather}
where
$\hat{T}=\mu+\Delta_{\perp}-(x^{2}+y^{2})+2|\varphi_{0}|^{2}$ and
$\omega$ are eigenvalues. Nonzero imaginary parts in $\omega$
imply the instability of the state $\varphi_{0}(\mathbf{r})$ with
$\max |\mathrm{Im}\,\omega|$ being the instability growth rate.
For radially symmetric vortices with $p=1$, azimuthal
perturbations turn out to be the most dangerous, and the problem
Eqs. (\ref{eigen1}) and (\ref{eigen2}) can be reduced to 1D
(radial) one and can then be easily solved with high accuracy
\cite{Malomed}. The situation, however, changes dramatically for
$p\neq 1$. In this case the radial symmetry is absent, and one
must solve the full 2D eigenvalue problem. The problem on a
$N\times N$ spatial grid implies a $2N^{2}\times 2N^{2}$ complex
nonsymmetric matrix and, for reasonable $N$ (say, $N>200$),
represents a formidable task. Instead, after inserting Eq.
(\ref{per}) into Eq. (\ref{main}), we solved the Cauchy problem
for the linearized equation
\begin{equation}
i\frac{\partial\varepsilon}{\partial
t}+\mu\varepsilon+\Delta_{\perp}\varepsilon-(x^{2}+y^{2})\varepsilon+
2|\varphi_{0}|^{2}\varepsilon+\varphi_{0}^{2}\varepsilon^{\ast}=0
\end{equation}
with some initial perturbations $\varepsilon$. The final results
are not sensitive to the specific form of $\varepsilon(x,y,0)$. If
the dynamics is unstable, the corresponding solutions
$\varepsilon(x,y,t)$, undergoing, generally speaking,
oscillations, grow exponentially in time and an estimate for the
growth rate $\gamma$ can be written as
\begin{equation}
\gamma=\frac{1}{2\Delta t}\ln\left\{\frac{P(t+\Delta
t)}{P(t)}\right\},
\end{equation}
where $P(t)=\int |\varepsilon|^{2}dxdy$, $t$ and $\Delta t$ are
assumed to be large enough.

\begin{figure}
\includegraphics[width=3.4in]{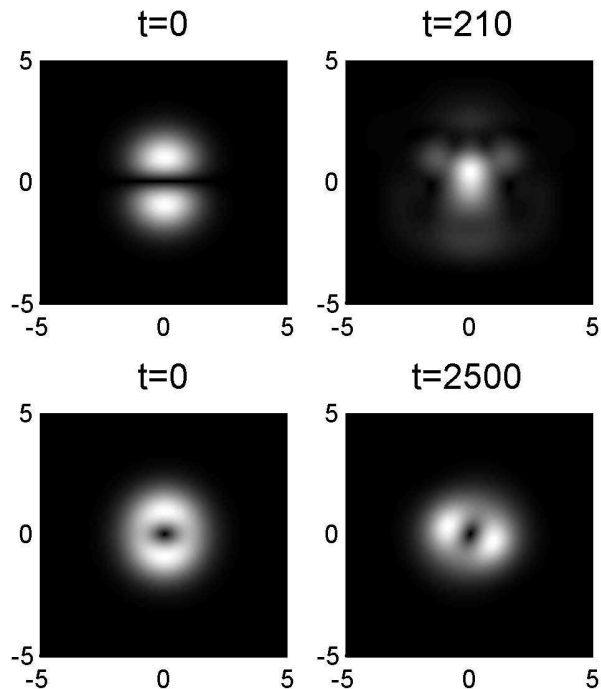}
\caption{\label{fig4} Top row -- unstable dynamics of the dipole
$p=0$ with $\mu=3.3$; bottom row -- stable evolution of the
azimuthon with $m=1$, $p=0.7$, and $\mu=2.7$.}
\end{figure}

In Fig. \ref{fig3} we plot the growth rates $\gamma$ as functions
of the chemical potential $\mu $ for $m=1$ and several different
values of $p$. The case $p=1$ corresponds to radially symmetric
vortices with the topological charge $m=1$ and was investigated in
detail in Ref. \cite{Malomed}. The corresponding curve in Fig.
\ref{fig3} coincides with that obtained in Ref.~\cite{Malomed}.
The linear stability analysis shows that for solutions with $m=1$
and $p\lesssim 0.7$, the growth rate of perturbations $\gamma\neq
0$ for all $\mu$ so that there is no the stability region. The
situation, however, changes when $p\gtrsim 0.7$. Under this, the
growth rate of perturbations $\gamma=0$ if the chemical potential
exceeds some critical value $\mu_{c}$. The stability window
appears and the azimuthons with $p\gtrsim 0.7$ and
$\mu_{c}<\mu<4$. The critical value $\mu_{c}$ monotonically
increases from $\mu_{c}\sim 2.45$ (for $p=0.7$) to $\mu_{c}\sim
2.6$ (for $p=1$). All solutions with $m=2$ turn out to be unstable
(for $p=1$ it was shown in Ref.~\cite{Malomed}).

To verify the results of the linear analysis, we solved
numerically the dynamical equation (\ref{main}) initialized with
our computed solutions with added Gaussian noise. The initial
condition was taken in the form $\varphi_{0}[1+\nu \xi(x,y)]$,
where $\varphi_{0}(x,y)$ is the numerically calculated solution,
$\xi(x,y)$ is the white gaussian noise with variance
$\sigma^{2}=1$ and the parameter of perturbation $\nu=0.005 \div
0.1$. The unstable dynamics of the nonrotating dipole ($p=0$) with
$\mu=3.3$ is illustrated in the top row of Fig.\ref{fig4}. Stable
evolution of the azimuthon with $p=0.7$ and $\mu=2.7$ (i. e. in
the region of the stability) is shown in the bottom row of
Fig.\ref{fig4}. The azimuthon cleans up itself from the noise and
survives over hundreds of the rotational periods.

In conclusion, we have presented nonrotating and rotating
(azimuthon) multisolitons in an effectively 2D ("pancake-shaped
configuration") Bose-Einstein condensate with attractive
interaction and parabolic trapping potential. We have performed a
linear stability analysis of these structures and demonstrated
that azimuthons with two intensity peaks (rotating dipoles) and
with not too small modulational depth can be stable if the number
of particles is below some critical value. The nonrotating
multisolitons (dipoles and all high-order multipoles) appear to be
unstable.

The author thanks Yu. A. Zaliznyak and A. I. Yakimenko for
discussions.

\end{document}